\documentclass[onecolumn,amssymb,showkeys,amsmath,nofootinbib,tightenlines,nobibnotes,aps,prl]{revtex4}

\usepackage[dvipsone]{graphicx}

\begin{document}

\title{Short-range effects and magnetization reversal in Co$_{80}$Fe$_{20}$ thin films: a MOKE magnetometry/domain imaging and AMR
study}


\author{J. M. Teixeira}\email{jmteixeira@fc.up.pt}
\author{R.F.A. Silva, J. Ventura, A. Pereira,
J. P. Araujo, M. Amado, F. Carpinteiro, J. B. Sousa}
\affiliation{IFIMUP and DFFCUP, Rua do Campo Alegre,678, 4169-007,
Porto, Portugal}
\author{S. Cardoso, R. Ferreira and
P. Freitas}\affiliation{INESC-MN and IST, Rua Alves Redol, 9-1,
1000-029 Lisbon, Portugal}

%
%



\begin{abstract}
A MOKE magnetometry unit simultaneously sensitive to both in-plane
magnetization components, based on an intensity differential
detection method, allows us to observe the uniaxial anisotropy
impressed during CoFe-deposition and to discriminate the
magnetization processes under a magnetic field parallel and
perpendicular to such axes. Our MOKE imaging unit, using a CCD
camera for Kerr effect domain visualization provides direct
evidence on the dominant M-processes, namely domain wall motion
and moment rotation. Further magnetic information was obtained by
AMR measurements due to the dependence of the electrical
resistivity on the short-range spin disorder and also on the angle
between the electrical current direction (I) and the spontaneous
magnetization ($\emph{\textbf{M}}_{S}$).
\end{abstract}\maketitle


\section{Introduction}
Ferromagnetic CoFe films with appropriated composition provide
highly spin-polarized 3d-electrons \cite{Wohlfarth} enhancing the
spin-dependent giant magnetoresistance. They are used in
spintronic devices such as magnetic tunnel junctions
\cite{Modera}, spin valves \cite{Spinvalves}, sensitive magnetic
sensors, high density read-heads \cite{readheads} and non-volatile
magnetic random access memories (MRAM) \cite{MRAMs}. Magnetic
switching and magnetization hysteresis is of particular relevance
for device functionality.

Here we present a detailed study based on the Anisotropic
Magnetoresistance (AMR), vectorial Magneto-Optical magnetometry
and domain imaging of a series of 200 \AA~Co$_{80}$Fe$_{20}$ films
with in-plane uniaxial anisotropy. A detailed description of these
techniques is given.

The easy and hard axis magnetization curves, M(H), are correlated
with different orientational processes, as inferred from MOKE
imaging. Further information was obtained from AMR measurements,
due to their dependence on short-range spin disorder and on the
angle between the electrical current direction and the spontaneous
magnetization ($\emph{\textbf{M}}_{S}$). Under transverse magnetic
fields, magnetic moment rotation dominate and lead to good
correlation between M, AMR and domain patterns. For longitudinal
fields the AMR(H) dependence cannot be correlated with M(H) since
significant AMR changes still occur when M is already saturated.
Such changes are discussed in terms of short-range spin disorder
effects.

\section{Experimental details}

The four probe technique was used for the AMR measurements, with
the electric current along the long axis and the applied magnetic
field parallel or at right angles to it. In ferromagnetic
3d-transition metals the electric resistivity depends on the angle
$\theta$ between the electrical current and the spontaneous
magnetization $\emph{\textbf{M}}_{S}$, through the so called
anisotropic magnetoresistive effect (Smit mechanism; see
\cite{AMR}):
\begin{equation}\label{Definição-de-R}
    \rho(\textbf{\emph{H}})=\rho_\perp+(\rho_\parallel-\rho_\perp)\cos^2\theta,
\end{equation}
where $\rho_{\perp}(\rho_{//})$ is the resistivity when
\emph{\textbf{M}} is saturated perpendicular (parallel) to the
electrical current. A magnetoresistive coefficient (at field
\emph{\textbf{H}}) is defined as:
\begin{equation}\label{Definição-de-AMR}
    \frac{\Delta\rho}{\rho}=\frac{\rho(\emph{\textbf{H}})-\rho(0)}{\rho(0)}.
\end{equation}
For a film with $\emph{\textbf{M}}_{S}$ in plane and starting with
a random demagnetized state one has
$\rho(0)=\frac{1}{2}\rho_{//}+\frac{1}{2}\rho_{\bot}$. If the film
has uniaxial anisotropy ($\pm\emph{\textbf{M}}_{S}$ domains) one
simply has $\rho(0)=\rho_{//}$. The so called anisotropic
magnetoresistance ratio (AMR), is given by \cite{AMR}:
\begin{equation}\label{Definição-de-AMR2}
AMR=\frac{\rho_{\parallel}-\rho_{\perp}} {\rho (0)}.
\end{equation}

\begin{figure}
    \centering
    \includegraphics[width=0.95\textwidth]{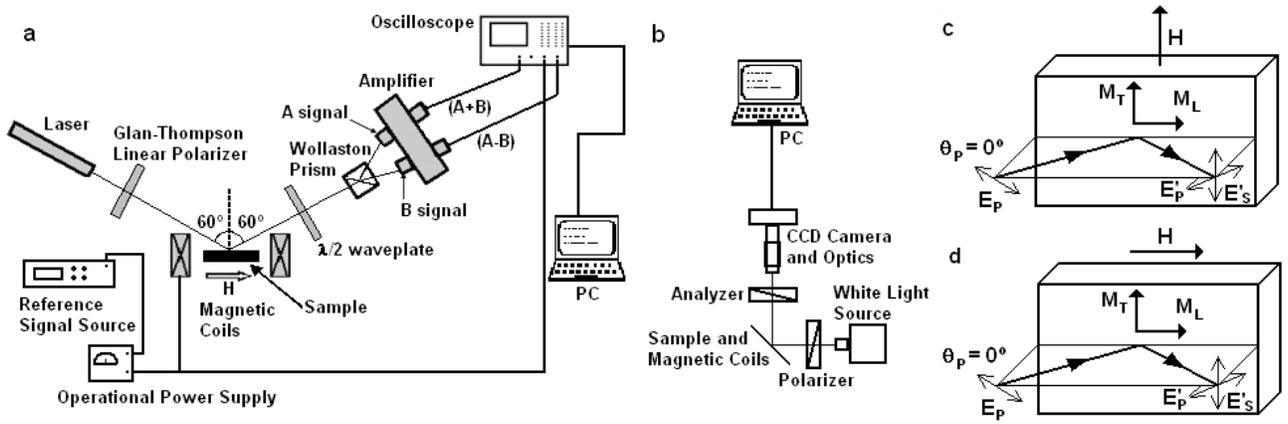}
    \caption{(a) MOKE magnetometry system used in this
    work. (b) Domain imaging system. (c)  Transverse MOKE geometry (d) Longitudinal MOKE
geometry.}\label{sistemasconjunto}
\end{figure}

Using the vectorial MOKE magnetometer depicted in Fig.
\ref{sistemasconjunto}a we measured the technical magnetization
$M(H)$ in the two common in-plane geometries
\cite{MOKEGeometries}: the \emph{transverse} geometry (Fig.
\ref{sistemasconjunto}c), with \textbf{\emph{H}} in the film plane
and perpendicular to the laser beam incident plane; and the
\emph{longitudinal} geometry (Fig. \ref{sistemasconjunto}d), in
which the in-plane field is parallel to the incident plane.
The light from a 5 mW He-Ne laser ($\lambda$ = 632.8 nm) passes
through a Glan-Thompson polarizer with an extinction ratio of
$1\times10^{-5}$, giving linearly polarized light parallel
(\emph{p}-polarization) to the plane of incidence. The sample is
located in the center of a pair of Helmholtz coils and in some
situations between the poles of an electromagnet. When convenient,
the (film) sample can be rotated in its plane. An operational
current supply produces a periodic (f=1Hz) magnetic field with
triangular shape. The laser light falls at $\sim60^\circ$ to the
film normal, and after reflection passes through a $\lambda/2$
dielectric plate, resulting in a $45^\circ$ rotation of the
polarization plane. The light is then split into two beams by a
Wollaston prism, corresponding to the two mutually orthogonal
(\emph{s} and \emph{p}) components of the polarized light (A and B
beams in Fig. \ref{sistemasconjunto}a), and the corresponding
intensities are measured by two independent photodiodes. Their
outputs are connected to a locally built amplifier, providing the
so called unbalanced and balanced MOKE signals. The unbalanced
signal (difference between the \emph{s} and \emph{p} components,
i.e. $A-B$ intensities) is proportional to the Kerr rotation and,
in a first approximation, it is a linear function of the
longitudinal (parallel to the plane of incidence) magnetization
component (M$_L$; Fig. \ref{sistemasconjunto}). The balanced
signal ($A+B$) is proportional to the square of the transversal
(perpendicular to the plane of incidence) magnetization component
(M$_T$; Fig. \ref{sistemasconjunto}). We can then simultaneously
measure the parallel and perpendicular technical magnetization
components, over the laser beam area, giving corresponding
averages over the magnetic moments. A 4-channel digital
oscilloscope is used to visualize and analyze the MOKE and the
magnetic field signals. Each magnetization curve is obtained by
averaging 128 successive hysteretic cycles.

A magnetic domain Kerr imaging system was also implemented to
visualize the magnetic domains (Fig. \ref{sistemasconjunto}b).
High resolution image digital processing enables us to obtain the
corresponding (averaged) technical magnetization. An halogen lamp
provides a non coherent light beam for this imaging system. This
light passes through a linear polarizer with the transmission axis
parallel to the plane of incidence (\emph{p}-polarization) and is
reflected at an angle of $\sim45^\circ$ with respect to the normal
of the sample. A periodic triangular shape magnetic field is
applied parallel to the film under study. An analyzer intercepts
the reflected beam with its transmission axis perpendicular to the
beam direction and is adjusted to transmit only one component of
the reflected light (the \emph{p}- or \emph{s}-component)
\cite{Flor}. A greyscale CCD camera with $\sim10 \mu$m of
resolution is used to acquire the magnetic domain images. Each
image is saved in a bitmap format with 8 bit of information and is
then subtracted from the magnetically saturated image to display
only the features associated with the magnetic behavior. The
magnetic field is produced by Helmholtz coils and acquired in an
oscilloscope. The AMR and Kerr imaging units can operate
simultaneously and are controlled (data acquisition and treatment)
with a Labview program.


\section{Experimental results}
Co$_{80}$Fe$_{20}$ 200 \AA~ thin rectangular (5 $\times$ 10 mm)
films were grown on glass substrates by ion-beam deposition
\cite{Ionbeam}, and subsequently annealed for 10 min at
280$^{\circ}$C. A magnetic field applied along the longitudinal
direction during deposition (3 kOe) always impressed a magnetic
easy axis.
\subsection{MOKE magnetometry and imaging}
\begin{figure}
    \centering
    \includegraphics[width=1\textwidth]{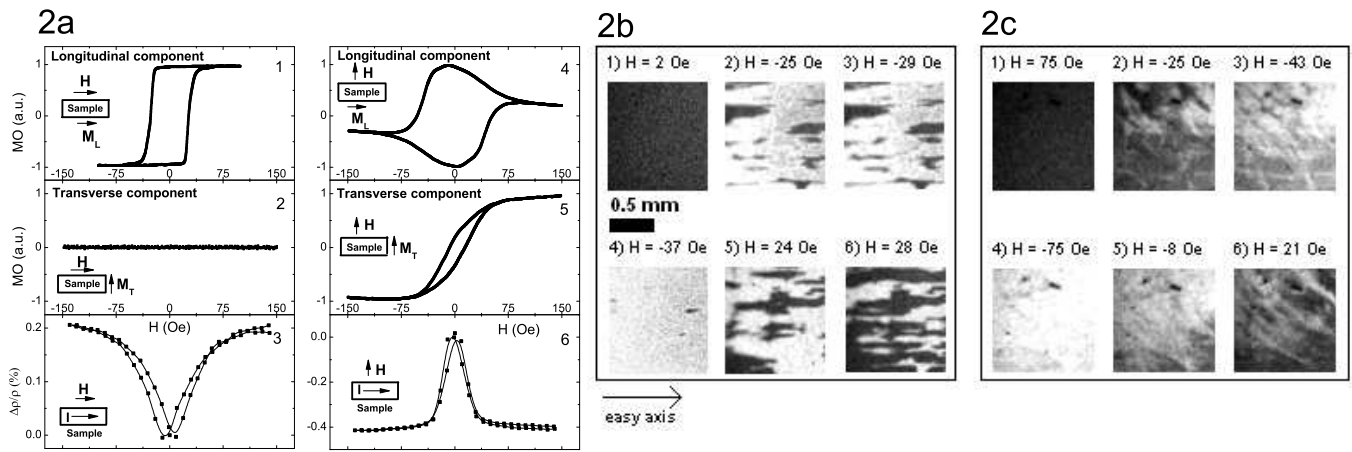}
    \caption{MOKE and AMR curves for the Co$_{80}$Fe$_{20}$ annealed film with
    \emph{\textbf{H}}
perpendicular [(a.4), (a.6) and (a.6)] and parallel [(a.1), (a.2)
and (a.3)] to the scattering plane. (b) Magnetic domains in CoFe
film along the forward (b.1)-(b.3) and backward (b.4)-(b.6)
branches of the $M_{L}$ loop depicted in a.1.
(\emph{\textbf{H}}$_a$ // easy axis) (c) Magnetic domains along
the forward (c.1)-(c.3) and backward (c.4)-(c.6) branches of the
$M_{T}$ loop depicted in a.5. (along hard
axis)}\label{Fig-upper-M}
\end{figure}

Typical longitudinal and transverse Kerr hysteresis loops for the
annealed CoFe film are presented in the four upper curves of Fig.
\ref{Fig-upper-M}, for an incident angle $\theta_{i} =
60^{\circ}$, using a p-polarized incident laser beam and covering
the two in-plane MOKE geometries (see Figs.
\ref{sistemasconjunto}c, d).

For \emph{\textbf{H}} along the easy axis, the longitudinal
component of the technical magnetization ($M_{L}$, Fig.
\ref{Fig-upper-M}a.1) exhibits a typical rectangular hysteretic
cycle, whereas the transverse component $M_{T}$ is zero (Fig.
\ref{Fig-upper-M}a.2; as expected for a soft ferromagnet
magnetized along its easy axis). The $M_{L}(H)$ cycle indicates a
magnetization process governed by sudden $180^{\circ}$-domain-wall
propagation \cite{Bozorth} (see below), leading to a coercive
field $H_{c} \sim 27$ Oe and a saturation field $H_{s} \sim 50$
Oe.

For \emph{\textbf{H}} perpendicular to the easy axis, Fig.
\ref{Fig-upper-M}a.5 shows a narrow magnetization $M_{T}(H)$
cycle, which is typical along the hard axis. A small coercive
field is observed ($H_{c} \sim 11$ Oe), whereas the saturation
field is in this case much higher, $H_{s} \sim 75$ Oe. On the
other hand, the behaviour of the longitudinal magnetization
component ($M_{L}$, Fig. \ref{Fig-upper-M}a.4) appears complex,
with magnetization reversal at $H_{c} \sim 50$ Oe and $H_{s} \sim
90$ Oe.



\subsection{AMR measurements}

The magnetoresistance $\Delta\rho/\rho$ vs $H$ curves for a
magnetic field parallel or perpendicular to the electrical current
\cite{AMR} are shown in Figs. \ref{Fig-upper-M}a.3 and
\ref{Fig-upper-M}a.6. The magnetoresistance is positive
($\Delta\rho/\rho \sim 0.2\%$) when the current and magnetic field
are parallel and negative ($\Delta\rho/\rho \sim -0.4\%$) when
they are perpendicular; an AMR ratio of $\sim$ 0.6\% results from
these data.

For the parallel configuration the $\Delta\rho/\rho$ curve
saturates much later ($H_{s} \sim 150$ Oe) than in the
perpendicular configuration ($H_{s} \sim 75$ Oe). In this context
one also notices that under longitudinal fields the AMR(H)
dependence does not correlate well with M(H) since important AMR
changes still occur when M is saturated. Under transverse magnetic
fields there is good agreement between M and AMR behaviour (see
below).





\section{Analysis of $M_{T}(H)$ and $M_{L}(H)$ behavior}

The rectangular easy-axis magnetization curve in Fig.
\ref{Fig-upper-M}a.1 clearly indicates that the magnetization
reversal is mainly due to sudden 180$^{\circ}$-domain-wall
formation and propagation, with no transverse magnetization
component ($M_T$; Fig.\ref{Fig-upper-M}a.2). This is directly
confirmed by the domain images obtained in the forward (Fig.
\ref{Fig-upper-M}b.1$\rightarrow$\ref{Fig-upper-M}b.3) and
backward branches (Figs.
\ref{Fig-upper-M}b.4$\rightarrow$\ref{Fig-upper-M}b.6) of the
$M_{L}(H)$ loop. These sequences clearly demonstrate the
nucleation of strip like magnetic domains, elongated along the
easy axis, coexistent with magnetic domains of opposite
magnetization (dark and bright regions), through 180$^\circ$
domain walls.

When the magnetic field is applied along the hard axis the growth
of the transverse magnetization is quite gradual, as revealed by
the intensity of the MOKE images (Figs.
\ref{Fig-upper-M}c.1$\rightarrow$\ref{Fig-upper-M}c.6). This
indicates the progressive rotation of the magnetization, as also
reflected in the $M_{L}(H)$ dependence (Fig.
\ref{Fig-upper-M}a.4); for a strictly coherent rotation one would
have $\tan \theta = M_{T}/M_{L}$.

\section{AMR versus M(H) behavior}

The magnetoresistance measurements show that this property is very
sensitive to the type of magnetic processes underlying the
magnetization reversal. Under transverse magnetic fields, when
magnetization rotation processes dominate, we have a good
correlation between the coercive and saturation fields extracted
from the $M(H)$ and $AMR(H)$ curves. This indicates that the
rotation processes, besides the magnetization, also dominate the
variation of resistivity under transverse magnetic fields, through
the Smit mechanism (eq.\ref{Definição-de-R}). For this transverse
field one still observes a longitudinal magnetization component,
but it should vanish for a sufficiently high field. However, due
to unavoidable small field misalignments with the hard axis,
ultimate $M_{L}$ vanishing is not strictly attained
(Fig.\ref{Fig-upper-M}a.4). It can be shown that the $M_{L}(H)$
component undergoes in the hysteresis cycle a 360$^\circ$ in-plane
rotation. At low fields the longitudinal magnetization component
displays a maximum amplitude and the magnetoresistance is very
small, since the magnetic moments are essentially along the easy
axis ($\theta \approx 0$ or $\pi$; M$_T$/M$_L$ $\ll$ 1) and,
according to eq.\ref{Definição-de-R},
$\rho(H)\sim\rho_{//}\sim\rho_0$ which leads to $\Delta\rho/\rho
\sim 0$.

For longitudinal fields the AMR(H) dependence cannot be correlated
with $M_{L}(H)$ since most of the AMR changes
(Fig.\ref{Fig-upper-M}a.3) occur when $M_{L}$ is already saturated
(long-range magnetic order; Fig. \ref{Fig-upper-M}a.1). Such
pronounced AMR changes are therefore due to short-range spin
disorder produced during the longitudinal magnetic switching,
which persists well above $H_{s}$ (as extracted from $M_{L}$).
Within De Gennes-Friedel model \cite{Gennes} the spin disorder
effects can be described by the expression $\rho(T,H) \simeq
\rho_{\infty}[1 - \frac{M^{2}_{S}(T,H)}{M^{2}_{s}(0)}\frac{S}{S+1}
+ \sum f(R_{ij})<\delta\vec{S_{i}}\cdot\delta\vec{S_{j}}>_{T,H}]$,
where $\rho_{\infty}$ is the magnetic resistivity in the
paramagnetic phase, $f(R_{ij})$ is the interference function for
the scattered electron wavefunctions and
$<\delta\vec{\vec{S}_{i}}\cdot\delta\vec{\vec{S}_{j}}>$ is the
correlation function for the spin fluctuations in different
lattice sites $\underline{i}$ and $\underline{j}$. For fully
uncorrelated spin fluctuations (PM state or mean field FM state)
one has $<\delta\vec{\vec{S}_{i}}\cdot\delta\vec{\vec{S}_{j}}> =
0$. If one includes some degree of spin-spin correlation in the
corresponding fluctuations, short range effects immediately affect
$\rho(T,H)$, through
$<\delta\vec{\vec{S}_{i}}\cdot\delta\vec{\vec{S}_{j}}>\neq 0$. In
this case, due to the small conduction electron wavelength
($\lambda_{F}$), only short-range effects are important in
resistivity, i.e. in distances up to $\lambda_{F}$. This means the
existence of short-range disorder effects in the nanometric range,
considerably above $H_{S}$.



Work supported in part by POCTI/0155, IST-2001-37334 NEXT MRAM and
POCTI/CTM/ 59318/2004 project. J. Ventura, R. Ferreira and S.
Cardoso are thankful for FCT grants (SFRH/BD/7028/2001,
SFRH/BD/6501/2001 and SFRH/BPD/7177/2001).


\begin{thebibliography}{12}

\bibitem{Wohlfarth}Wohlfarth, \emph{Ferromagnetic
Materials; vol. 2}, North-Holland Company New York (1980).
\bibitem{Modera}J. S. Moodera, L. R. Kinder, T. M. Wong and R. Meservey, \emph{Phys.
Rev. Lett}. {\bf 74}, 3273 (1995).
\bibitem{Spinvalves}J. B. Sousa \emph{et al.}, \emph{J. Appl. Phys.} {\bf 91}, 5321 (2002).
\bibitem{readheads}D. Song, J. Nowak, R. Larson, P. Kolbo, R. Chellew,
\emph{IEEE Trans. Magn.} \textbf{36} 2545 (2000).
\bibitem{MRAMs}S. Tehrani\emph{et al.}, \emph{IEEE
Trans. Magn.} \textbf{36} 2752 (2000).
\bibitem{Ionbeam}V. Gehanno \emph{et al.}, \emph{IEEE Trans. Magn.} {\bf 35},  4361 (1999).
\bibitem{AMR}T. R. McGuire and R. I. Potter, \emph{IEEE Trans. Magn.} {\bf
11}, 1018 (1975).
\bibitem{MOKEGeometries}C. Daboo \emph{et al.}, \emph{Phys.
Rev. B}. {\bf 47}, 11852 (1993).
\bibitem{Flor}J. M. Florczak and E. Dan Dahlberg, \emph{J. Appl. Phys.} {\bf 67}, 7520 (1990).
\bibitem{Bozorth}Richard M. Bozorth, \emph{Ferromagnetism}, IEEE Press
(1951).
\bibitem{AlexHubert}Alex Hubert and Rudolf Sch\"{a}fer, \emph{Magnetic
Domains}, Springer (1998).
\bibitem{Gennes}P.G. De Gennes and J. Friedel, \emph{J. of Phys. Chem.
Solids.} {\bf 4}, 71 (1958).


\end{thebibliography}
\end{document}